Conditional beta and uncertainty factor in the cryptocurrency pricing model

Nguyen Quoc Khanh

Brisbane, Queensland

September 2020

# 1. Introduction

Cryptocurrencies are increasingly becoming a popular and captivating investment in portfolios. Its returns are incredibly high but also come with high risks and volatility compared to traditional investments. Many recent studies have committed themselves to build pricing asset models (Liu et al., 2020; Shen et al., 2019). Most of these valuation models are based partly on traditional assets pricing models such as the single-factor CAPM model (Sharpe, 1964), the three-factor model (Fama and French, 1993), the four factors with additional momentum (Cahart, 1997), and the five-factor model (Fama and French, 2005).

Although cryptocurrencies have some similar characteristics to stocks (Chan et al., 2017; Gkillas and Katsiampa, 2018; Phillip et al., 2019), the significant difference is that its price is not based on any fundamental value. In addition to the usual factors that affect prices, it is also strongly influenced by investor sentiment. One of the factors influencing investor sentiment is the uncertainty. Baker et al. (2016) indicated that the uncertainty significantly affects stock price volatility and investment in some sectors. So we will incorporate the uncertainty into the cryptocurrency pricing model as conditional information to examine the impact of factors on price. There has been no previous study that has included this factor in the pricing model of cryptocurrencies and stocks as well.

Most pricing models assume that the beta is fixed. However, while allowing the beta changes over time and depends on macroeconomic variables and stock characteristics, pricing models better capture the impact of size, value effect and past returns (Avramov and Chordia, 2006). Ho and Hung (2009) also used beta conditional on investor sentiment, default spread and stock's characteristics and then found that investment sentiment also helped to capture the anomalies impact in the stock pricing model better. We will also apply conditional beta that depends on the uncertainties, cryptocurrency's characteristics and Bitcoin's return.

Moreover, the study will, in turn, examine the factors that influence prices in the previous models including the market risk premium in CAPM (Sharpe, 1964), the size and value risk factors (Fama and French, 1993), the momentum factor (Carhart, 1997), and the liquidity factor (Pastor and Stambaugh, 2003).

# 2. Objectives and research questions

The objective of the research is to build a pricing model of cryptocurrencies by using conditional beta and uncertainty factor. It will answer the following research questions. *Firstly*, what is the difference in pricing model when we use the typical unconditional beta and the conditional beta? *Secondly,* when we add uncertainty factor into the pricing models, do the anomalies explain for return better? *Thirdly*, which anomalies are essential to explain returns for cryptocurrencies?

## 3. Significance

This research contributes to literature at the following points. Firstly, it is the first study to assess cryptocurrencies with the conditional beta, compared with prior studies based on unconditional beta or fixed beta. It is a new approach to building a pricing model for cryptocurrencies. Therefore, we expect that the use of conditional beta will increase the explanatory ability of factors in previous pricing models. Secondly, this research is also a pioneer in placing the uncertainty factor into the cryptocurrency pricing model. Earlier studies on cryptocurrency pricing have ignored this factor. However, it is a significant factor in the valuation of cryptocurrencies because the uncertainty leads to the investor sentiment and affects prices.

## 4. Methodology

This study is based on the two-pass regression framework of Avramov and Chordia (2006), which is summarized as in equation (1). Specifically, in the first-pass time-series regressions as (2), it regresses the daily cryptocurrency's risk premium on the risk factors ($F_t$) based on previous pricing models and conditional beta ($\beta$). In the second-pass cross-sectional equation as (3), the study runs regression of the risk-adjusted returns ($R^*_{jt}$), which is the sum of the intercept and the residual in (2), on the cryptocurrency's characteristics including size, liquidity, and momentum.

$$R^*_{jt} \equiv R_{jt} - [\ R_{Ft} + \beta\ (\theta;\ S_{t-1},\ X_{jt-1}\ )'\ F_t\ ] = c_{ot} + c_t\ Z_{jt-1} + e_{jt} \quad (1)$$

$$R_{it} - R_{Ft} = \alpha_j + [\beta\ (\theta;\ U_{t-1},\ R_{t-1},\ X_{jt-1}\ )'\ F_t\ ] + u_{jt} \quad (2)$$

$$R^*_{jt} = c_{ot} + c_t\ Z_{jt-1} + e_{jt} \quad (3)$$

Where $R_{jt}$ is the return of cryptocurrency j at time t, and $R_{Ft}$ is the risk- free rate. This research regresses the risk-adjusted return on the cryptocurrency's characteristic variables in (3) based on the approach of Brennan, Chordia, and Subrahmanyam (1998) and Shanken (1992) that suggested that this method would avoid finite sample bias attributable to errors when regressing factors in the first-pass time-series equation (2). So this research expects the low explanation ability of anomalies on risk- adjusted return in (3).

In this study, the macroeconomic variables ($S_{t-1}$) in (1) will be replaced by the uncertainty factor ($U_{t-1}$) and Bitcoin return ($R_{t-1}$) as in (2). θ are the parameters that capture the relationship between β and the uncertainty, Bitcoin return, and the cryptocurrency's characteristics ($X_{jt-1}$). $F_t$ is a vector representing risk factors that depends on the previous pricing models. $Z_{jt-1}$ is a vector of anomalies, including size, liquidity, and momentum of a cryptocurrency j. For specification, the conditional beta (β) is calculated as follows

$$\beta_{jt-1} = \beta_{j1} + \beta_{j2} U_{t-1} + \beta_{j3} R_{jt-1} + (\beta_{j4} + \beta_{j5} U_{t-1} + \beta_{j6} R_{jt-1}) SIZE_{jt-1}$$
$$+ (\beta_{j7} + \beta_{j8} U_{t-1} + \beta_{j9} R_{jt-1})_m C_{m\ jt-1} \quad (4)$$

Where SIZE is the market capitalization of cryptocurrency j, and $C_m$ are the other determinants. This beta is then placed to regress equation (2). The risk factors in vector F depend on the previous models that this study will examine. For example, if it is the single-factor CAPM model (Sharpe, 1964), then F has one variable- market risk premium (Rm). Similarly, F would consist of 3 variables, market risk premium, size and value factor, when we use the 3-factor model of Fama and French (1993). This study will, in turn, examine the previous typical pricing models, including CAPM (Sharpe, 1964), 3-factor model (Fama and French, 1993), 4-factor model (Carhart, 1997) with the added momentum factor based on the 3-factor model, the 4-factor model with added liquidity factor (Pastor and Stambaugh, 2003) based on the 3-factor model, and the model using all the above factors. To illustrate, if we firstly examine the CAPM model with only market risk premium (Rm), equation (2) with conditional beta will become

$$R_{it} - R_{Ft} = \alpha_j + Rm\ [\beta_{j1} + \beta_{j2} U_{t-1} + \beta_{j3} R_{jt-1} + (\beta_{j4} + \beta_{j5} U_{t-1} + \beta_{j6} R_{jt-1}) SIZE_{jt-1}$$
$$+ (\beta_{j7} + \beta_{j8} U_{t-1} + \beta_{j9} R_{jt-1})_m C_{m\ jt-1}] + u_{jt} \quad (5)$$

The risk-adjusted returns $R^*_{jt}$ will be calculated as the sum of the intercept $a_j$ and residual $u_{jt}$ in (5) and be used to regress (3) that is $R^*_{jt} = c_{ot} + c_t Z_{jt-1} + e_{jt}$. Where $Z_{jt-1}$ is a vector of anomalies, including size, liquidity, and momentum of a cryptocurrency j. Likewise, this study will investigate all of the models as mentioned above.

## 5. Expected Results

In most of the pricing models examined, using the conditional beta will better capture the impact of factors on cryptocurrency return than while using the unconditional beta. So when using the conditional beta in (2), the adjusted $R^2$ in (3) is expected to be lower, and the anomalies ($Z_{jt-1}$) have an insignificant impact on the risk-adjusted returns ($R^*_{jt}$.) Besides, including the uncertainty factor into the pricing models will capture the effects of anomalies on cryptocurrency risk-adjusted returns.

## 6. Data

This study will use data of cryptocurrencies in the Coinmarketcap website from the beginning of 2016 to the present. The data extracted from the website is detailed about the daily prices, volumes, and market capitalization of cryptocurrencies. We use the closing price of one day and the next one to calculate the daily return. The cryptocurrencies chosen for analysis are those in the top 200 and have existed for at least one year. The risk-free rate is one-month U.S. Treasury bill rate. However, research will also try to alter it with the Bitcoin return as a robustness check because investors expect a higher return on alt-coins than Bitcoin's. Regarding the index for the uncertainty factor, the study uses the economic policy uncertainty (EPU) index, proposed by Baker et al. (2016) based on newspaper coverage frequency. This index correlated significantly with stock price volatility and some macroeconomic variables. It is statistical daily and can be used conveniently to analyse the daily return of cryptocurrencies accordingly.